\documentclass[letterpaper,12pt]{emulateapj}
\usepackage{color}
\usepackage{longtable}
\usepackage{dashrule}
\usepackage{hyperref}
\begin{document}
\title{Early-time VLA observations and broad-band afterglow analysis\\ of
the Fermi-LAT detected GRB\,130907A}
\author{P\'eter Veres \altaffilmark{1,*}, Alessandra Corsi \altaffilmark{2},
Dale A. Frail \altaffilmark{3}, S. Bradley Cenko \altaffilmark{4,5}, Daniel A.
Perley \altaffilmark{6}
}
\altaffiltext{1}{
The George Washington University, Department of Physics, 725 21st St, NW,
Washington, DC 20052, USA
}
\altaffiltext{2}{
Texas Tech University, Department of Physics, Box 41051, Lubbock, TX 79409-1051, USA
}
\altaffiltext{3}{
National Radio Astronomy Observatory, P.O. Box O, Socorro, NM 87801, USA
}
\altaffiltext{4}{
Astrophysics Science Division, NASA Goddard Space Flight Center, Mail Code 661,
Greenbelt, MD 20771, USA
}
\altaffiltext{5}{
Joint Space-Science Institute, University of Maryland, College Park, MD 20742, USA
}
\altaffiltext{6}{
Department of Astronomy, California Institute of Technology, MC 249-17, 1200
East California Blvd, Pasadena CA 91125, USA
}

\altaffiltext{*}{{ Email: \email{peter.veres@uah.edu}, currently at: Center for
Space Plasma and Aeronomic Research (CSPAR), University of Alabama in
Huntsville, Huntsville, AL 35899, USA} }
\date{\today} 
\def\ve{\varepsilon}
\def\gbm{{\it GBM }}
\def\lat{{\it LAT }}
\def\fermi{{\it Fermi }}
\def\swift{{\it Swift }}
\newcommand{\Mesz}{{M\'esz\'aros}}
\def\mathnew{\mathsurround=0pt}
\def\simov#1#2{\lower .5pt\vbox{\baselineskip0pt \lineskip-.5pt
      \ialign{$\mathnew#1\hfil##\hfil$\crcr#2\crcr\sim\crcr}}}
\def\simg{\mathrel{\mathpalette\simov >}}
\def\siml{\mathrel{\mathpalette\simov <}}
\def\beq{\begin{equation}}
\def\enq{\end{equation}}
\def\bea{\begin{eqnarray}}
\def\ena{\end{eqnarray}}
\def\bitm{\bibitem}
\def\msun{M_\odot}
\def\L54{L_{54}}
\def\E55{E_{55}}
\def\et3{\eta_3}
\def\th1{\theta_{-1}}
\def\r07{r_{0,7}}
\def\x05{x_{0.5}}
\def\et600{\eta_{600}}
\def\et3{\eta_3}
\def\rph{r_{ph}}
\def\vareps{\varepsilon}
\def\fflunit{\hbox{~erg cm}^{-2}~\hbox{s}^{-1}}
\def\eps{\epsilon}
\def\ve{\varepsilon}
\def\muh{\hat{\mu}}
\def\cm{\hbox{~cm}}
\def\kpc{\hbox{~kpc}}
\def\Mpc{\hbox{~Mpc}}
\def\km{\hbox{~km}}
\def\s{\hbox{~s}}
\def\gev{\hbox{~GeV}}
\def\Jy{\hbox{~Jy}}
\def\TeV{\hbox{~TeV}}
\def\GeV{\hbox{~GeV}}
\def\MeV{\hbox{~MeV}}
\def\kev{\hbox{~keV}}
\def\keV{\hbox{~keV}}
\def\eV{\hbox{~eV}}
\def\G{\hbox{~G}}
\def\Hz{\hbox{~Hz}}
\def\GHz{\hbox{~GHz}}
\def\ghz{\hbox{~GHz}}
\def\erg{\hbox{~erg}}
\def\s{{\hbox{~s}}
\def\cm2{\hbox{~cm}^2}}
\def\para{\parallel}
\def\Fl{\mathcal{F}}
\defcitealias{Veres+12magnetic}{VM12}
\defcitealias{Sari+00refresh}{SM00}
\begin{abstract}
We present multi-wavelength observations of the  hyper-energetic gamma-ray
burst (GRB) 130907A, a \swift-discovered burst with early radio observations
starting at $\approx 4$\,hr after the $\gamma$-ray trigger. GRB\,130907A was
also detected by the \textit{Fermi}/LAT instrument and, at late times, showed a
strong spectral evolution in X-rays.  We focus on the early-time radio
observations, especially at $>10\,\GHz$, to attempt identifying reverse shock
signatures. While our radio follow-up of GRB\,130907A ranks among the earliest
observations of a GRB with the Karl G. Jansky Very Large Array (VLA), we did
not see { an unambiguous } signature of a reverse shock.  { While a model
with both reverse and forward-shock can correctly describe the observations,
the data is not constraining enough to decide upon the presence of the
reverse-shock component.} We model the broad-band data using a simple
forward-shock synchrotron scenario { with} a transition from a wind
environment to a constant density interstellar medium (ISM) in order to account
for the observed features. { Within the confines of this model, we} also
derive the underlying physical parameters of the fireball, which are within
typical ranges except for the wind density parameter ($A_*$), which is higher
than those for bursts with wind-ISM transition, but typical for the general
population of bursts. We note the importance of early-time radio observations
of the afterglow (and of well sampled light curves) to unambiguously identify
the potential contribution of the reverse shock.  \end{abstract}

\section{Introduction}
\label{sec:intro}
Gamma-ray bursts' afterglows still pose some fundamental unanswered questions.
The processes giving rise to prompt and early-afterglow emission at optical and
radio frequencies are among the least well understood. Early-time optical
\citep{Akerlof99}  and radio \citep{kulkarni99} flashes were first discovered
in GRB\,990123 and attributed to reverse-shock emission
\citep[e.g.][]{Meszaros+99rev, Sari+99optflash2, Corsi+05, Urata+14rsssc}. But,
later on, fast robotic telescopes did not find evidence for early optical
flashes in the expected numbers \citep{Melandri+08opt}.  It has been suggested
that the lack of early optical emission may be due to the fact that the reverse
shock peaks below the optical range, at mm or cm wavelengths \citep{kulkarni99,
Chandra2012, Laskar2013, Perley2014}.  Another possibility is that the reverse
shock is entirely suppressed by e.g. a high degree of magnetization of the
ejecta \citep{Zhang+05rsmag}.

Here, we present early-time radio observations of GRB\,130907A, together with
observations at other wavelengths.  Our radio follow-up of GRB\,130907A ranks
among the earliest observations of a GRB with the VLA. However, our data do not
show a clear reverse shock signature.  Besides an early-time radio follow-up
and a self-absorbed radio spectrum, the other interesting features of this
burst consist of an early-time \textit{Fermi}/LAT detection and a significant
late-time spectral evolution in the X-ray band.

Our paper is organized as follows. In Section\,\ref{sec:obs} we present the
observational data for this burst  and discuss the spectral and temporal
properties of GRB\,130907A. In Section\,\ref{sec:model} we provide a
theoretical interpretation for the broad-band data, and conclude in
Section\,\ref{sec:conclusion}. In this paper, we use the $F_\nu\propto
t^{-\alpha}\nu^{-\beta}$ notation ($\alpha$ is the temporal index and $\beta$
is the spectral index) and $Q=10^x Q_x$ for any physical quantity $Q$ in cgs
units (unless otherwise stated). 

\section{Observations and data analysis}
\label{sec:obs}
\subsection{Gamma-rays }
GRB\,130907A \citep{2013GCN..15183}, was discovered by the BAT instrument
aboard the \swift satellite \citep{gehrels04} at 21:41:13 UT.  It was also
detected by the \textit{Fermi}/LAT \citep{2013GCN..15196}, Konus WIND
\citep{2013GCN..15203}, and various ground-based observatories at longer
wavelengths \citep[e.g.][]{2013GCN..15220, 2013GCN..15200}.  With a redshift of
$z=1.238$, this GRB occurred at a luminosity distance of $D_L=2.7\times
10^{28}\cm $ \citep{2013GCN..15187}, calculated using the following
cosmological parameters: $\Omega_m=0.27$, $\Omega_\Lambda=0.73$ and $H_0=73 \km
\s^{-1} \Mpc^{-1}$.  The isotropically emitted energy is $E_{\rm iso }\sim 3.0
\times 10^{54} \erg$ { (calculated from 1 keV to 10 MeV in the local frame)}.
The jet opening angle ($\theta_j\gtrsim12^\circ$, see Section \ref{sec:xbrk})
corrected energy is $E_{\rm jet}\gtrsim3\times10^{52}$ erg, which makes this
burst part of the hyper-energetic class of GRBs \citep{Cenko+11hyper}.
Moreover, at 18\,ks after the $\gamma$-ray trigger, the \textit{Fermi}/LAT
detected one of the highest-energy (55\,GeV) photons ever observed in a GRB
\citep{2013GCN..15196}.  We refer the reader to \citet{Tang+14ic} for more
details about the LAT flux measurements. 

\begin{figure}[htbp]
\begin{center}
\includegraphics[width=0.999\columnwidth,angle=0]{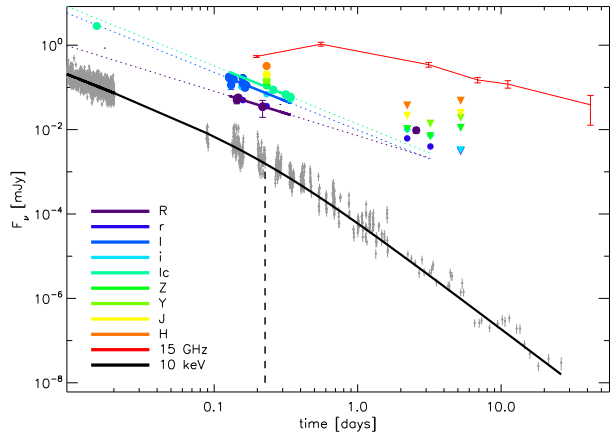}
\caption{Optical, radio, and X-ray flux measurements of GRB\,130907A. The fit to
the optical data is only for the intervals marked with thick continuous lines, the
dotted lines are extrapolations. Triangles mark upper limits. The vertical line
marks the temporal break in the
X-ray fit.}
\label{fig:Xlc} 
\end{center}
\end{figure}

\subsection{X-rays}
\label{sec:xlc}
\begin{figure}[htbp]
\begin{center}
\includegraphics[width=0.999\columnwidth,angle=0]{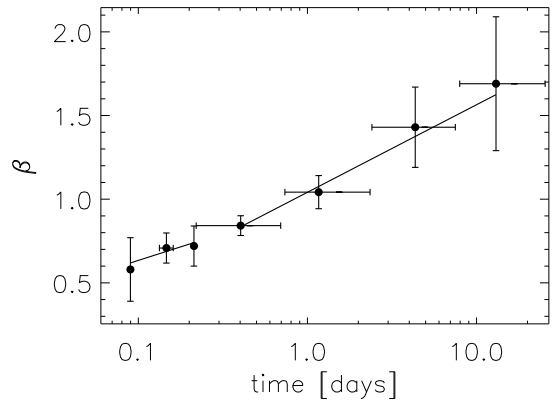}
\caption{Evolution of the X-ray spectral index with time. Spectral indices were
obtained
using the XRT spectral repository (see Section\,\ref{sec:xlc}). }
\label{fig:spidx} 
\end{center}
\end{figure}

X-ray measurements by \swift/XRT of GRB\,130907A started at $\approx 56$\,s
after the trigger (during the burst prompt $\gamma$-ray emission phase) and
lasted until $\approx 26$\,d after the burst. The light curve is overall
declining with an easily identifiable break around 0.2\,d since trigger
(Figure\,\ref{fig:Xlc}), and a strong spectral evolution at late times
(Figure\,\ref{fig:spidx}). Due to this spectral evolution, we used a dynamic
count-to-flux density conversion and derived an accurate $10 \keV$ light curve
using the {\it burst analyser} \citep{evans07,Evans+10analyser}.

The 10\,keV light curve of GRB\,130907A shows a clear break followed by a
steepening of the temporal decay index (Figure\,\ref{fig:Xlc}). In what
follows, for our analysis of the afterglow, we discard all observations at
$t<300 \s$ due to possible contribution of the prompt emission.  By fitting the
X-ray light curve with a smoothly broken power-law of the form $F_\nu=A [
(t/t_{\rm break})^{\alpha_1} +(t/t_{\rm break})^{\alpha_2}]^{-1}$
\citep{Beuermann+99sbpl} we get $t_{\rm break} = 0.23\pm0.02 $ days and
indices: $\alpha_1=1.32\pm 0.02$ and $\alpha_2=2.57\pm 0.05$, respectively
before and after the break. { We did not attempt to find the parameter
responsible for the smoothness of the break and fixed it to the nominal value
of 1 \citep[the $s$ parameter in][]{Beuermann+99sbpl}. This could tentatively
explain why the fit underestimated the points close to the break.} 

We have obtained the spectral data from the XRT repository's spectral
tools\footnote{\url{http://www.swift.ac.uk/xrt_spectra}}.  The spectral index
before the temporal break is unusually hard,  $\beta_{\rm X,early}=0.69\pm0.06$
with no significant evolution.  A unique feature of the X-ray afterglow is the
spectral evolution starting with the light curve break, from
$\beta_X=0.8\pm0.1$ at early times to $\beta_X=1.7\pm0.4$ at later times (see
Figure \ref{fig:spidx}).  A linear fit (in $\log t-\beta$) to the first three
points gives a slope of $0.32\pm0.58$, consistent with no spectral evolution,
while for the last four points the slope is $0.52\pm 0.16$ indicative of an
evolving spectral index.  The average spectral index after the break is
$\beta_{\rm X,late}= 0.96\pm 0.05$.  The X-ray absorbing column of the host
galaxy is: $N_H=(9.8\pm1.1)\times10^{21} \cm^{-2}$.

We finally note that the light curve integrated for the entire energy interval
of XRT (0.3-10 keV) has a different behavior than the flux density light curve
plotted in Figure \ref{fig:Xlc}.  { Indeed, the automatic fitting routine
\citep{Evans+09auto} yields a broken power-law fit with three breaks for the
integrated light curve (see
\url{http://www.swift.ac.uk/xrt_live_cat/00569992/}). On the other hand, a fit
with a smoothly broken power-law function (similar to the one we fitted to the
flux density light curve) does not constrain the break time, while a simple
power-law fit gives a slope of $\alpha=1.510\pm 0.003$. The latter is in strong
contrast with the steep slope found for the late-time flux density (Figure 1),
and it is indicative of a changing X-ray spectrum at late times.
}

\subsection{Optical}
\label{sec:optdata}
We gathered the optical observations of GRB\,130907A from the public GCN
bulletins (see Table\,\ref{tab:optical}).  In case of Skynet observations, where
the data points were reported on figures, we obtained the numerical flux values
by digitizing these figures.

We correct the optical data of GRB\,130907A for Galactic absorption ($A_V=0.03$
mag) in the direction of the burst using the maps of \citet{Schlafly+11galext}.
We find that the optical spectral index at the time of the first radio
observation is $\approx 2.5$, suggesting that the host galaxy strongly absorbs
the optical flux.  We fit the I, Ic and R filter measurements from $0.12 \,$d
to $0.35 \,$d to obtain the temporal decay index for this time interval (Figure
\ref{fig:Xlc}), and we get $\alpha_I=1.37 \pm0.38$ (from 7 measurements;
Table\,\ref{tab:optical}), $\alpha_{Ic}= 1.38\pm 0.59$ (from 3 measurements;
Table\,\ref{tab:optical}), and $\alpha_R=1.05 \pm1.06 $ (from 3 measurements;
Table\,\ref{tab:optical}).  The other optical observations, at earlier and
later times, are too sparse to derive secure spectral or temporal information.
We note, however, that late-time measurements seem to lie above the
extrapolation from the temporal decay derived between $0.12 \,$d to $0.35 \,$d,
thus suggesting a late-time flattening (Figure \ref{fig:Xlc}). { A foreground
galaxy (SDSS J142333.95+453626.2 with photometric redshift $z=0.6\pm0.3$) at
a distance of $\approx 0.5''$ from GRB 130907A has been identified by
\citet{2013GCN..15192} with $i$ and $r$ magnitudes comparable to that of the
afterglow at $\approx 3$ days after the trigger. This galaxy could tentatively
explain the flattening of the optical afterglow at these late times.  On the
other hand, \citet{2013GCN..15223} have limited the contribution of galaxy flux
in $r$- and $i$ bands to $>$ 22.6 mag. }

\begin{figure}[htbp]
\begin{center}
\includegraphics[width=0.999\columnwidth,angle=0]{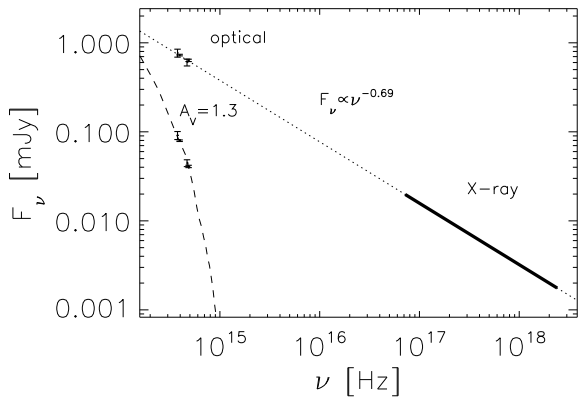}
\caption{Illustration of the host galaxy extinction on the optical measurements
at Epoch I. The thick line represents the measured X-ray spectrum, the dotted
line is its extrapolation. Dashed lines show the absorbed spectral energy
distribution which fits the observations (measurements atop the dashed curve).
The points on the dotted line are the de-absorbed data points.}
\label{fig:extcorr} 
\end{center}
\end{figure}

To account for the extinction (reddening) in the host galaxy, we use the
absorption curves of \cite{Gordon+03hostext}.  The relation between the
absorbed and unabsorbed flux is: $F_\nu^{\rm obs} = F_\nu^{\rm unabs} \times
10^{-0.4 A(\nu)} $. Here, $A(\nu)$ is the absorption curve, characterized by a
$V$-band value, $A_V$, which is a free parameter, and a shape which is usually
taken to be similar to the LMC, SMC or the Milky Way.  Due to the small number
of optical observations, we assume the SMC extinction curve, which is most
commonly used for GRB afterglow studies \citep{Schady+10dust}. We have the best
spectral coverage in the optical band around the time of Epoch I { ($t=0.193$
days), where we have extrapolated the available filters (R, Rc, I, Ic) from
$\sim 0.1$ to $\sim 0.4$ days}.  By further assuming that the optical and X-ray
spectral indices are the same ($\beta=0.69$; at this early time there is no
significant spectral evolution in X-rays, and the optical and X-ray temporal
slopes are consistent with being the same within the large uncertainties) we
find $A_V\approx 1.3\pm 0.1$ in the frame of the host galaxy (see Figure
\ref{fig:extcorr}). 

{ The above value of $A_V$ is derived by fitting a power law with the same
spectral index as the X-ray measurements to the unabsorbed optical points.
Indeed, we can exclude a spectral break between the optical and X-ray regimes:
If there was a break, the optical spectral index would be $\beta=-1/3$ which,
after correction for extinction, would imply a true optical flux incompatible
with  the X-ray data. } 

Using the X-ray spectral analysis described in Section\,\ref{sec:xlc}, we
estimate $N_H/A_V\approx 7.5\times 10^{21} \cm^{-2}$. This value is on the
lower side, but still consistent with, the $N_H/A_V$ distribution reported in
\cite{Schady+10dust}.  Generally speaking, GRB host galaxies have
systematically larger $N_H/A_V$ ratios compared to the Magellanic clouds and
the Milky Way \citep{Schady+10dust}, and this effect is at least partly
intrinsic to the host galaxies.  Bearing in mind the uncertainties on $A_V$ due
to the small number of optical measurements,  the  $N_H/A_V$ value of the
host galaxy of GRB\,130907A is more like the SMC than for most GRB sightlines.

\subsection{Radio}
\label{sec:radiolc}
Radio observations of GRB\,130907A were performed with the VLA\footnote{The
National Radio Astronomy Observatory is a facility of the National Science
Foundation operated under cooperative agreement by Associated Universities,
Inc.; \url{http://www.nrao.edu/index.php/about/facilities/vlaevla}}
\citep[][]{Perley2009} in its CnB and B configurations, under our Target of
Opportunity programs\footnote{VLA/13A-430 - PI: A.\,Corsi; VLA/S50386 - PI:
S.B.\,Cenko}. 

\begin{figure}[htbp]
\begin{center}
\includegraphics[width=0.399\columnwidth,angle=0]{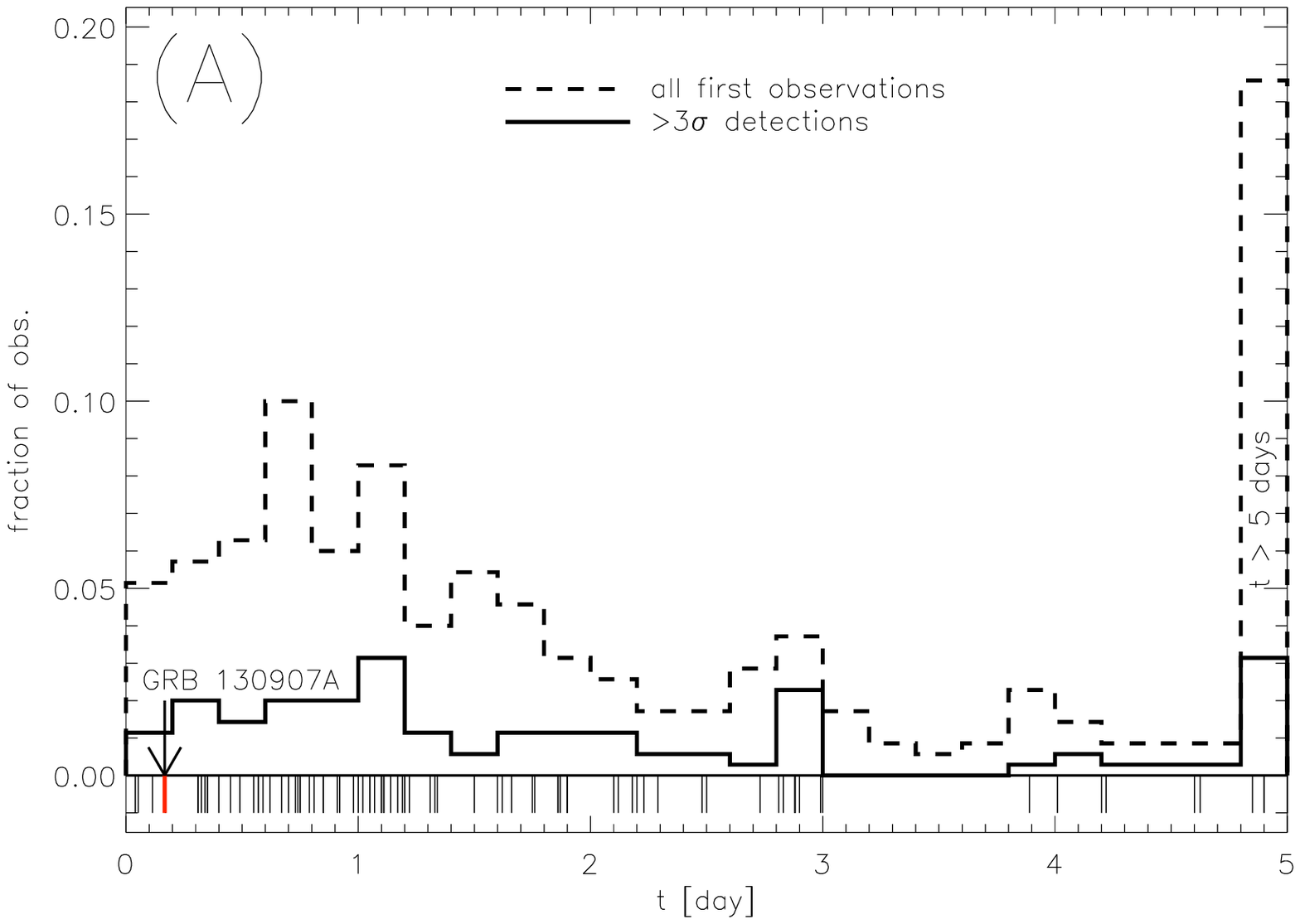}
\includegraphics[width=0.399\columnwidth,angle=0]{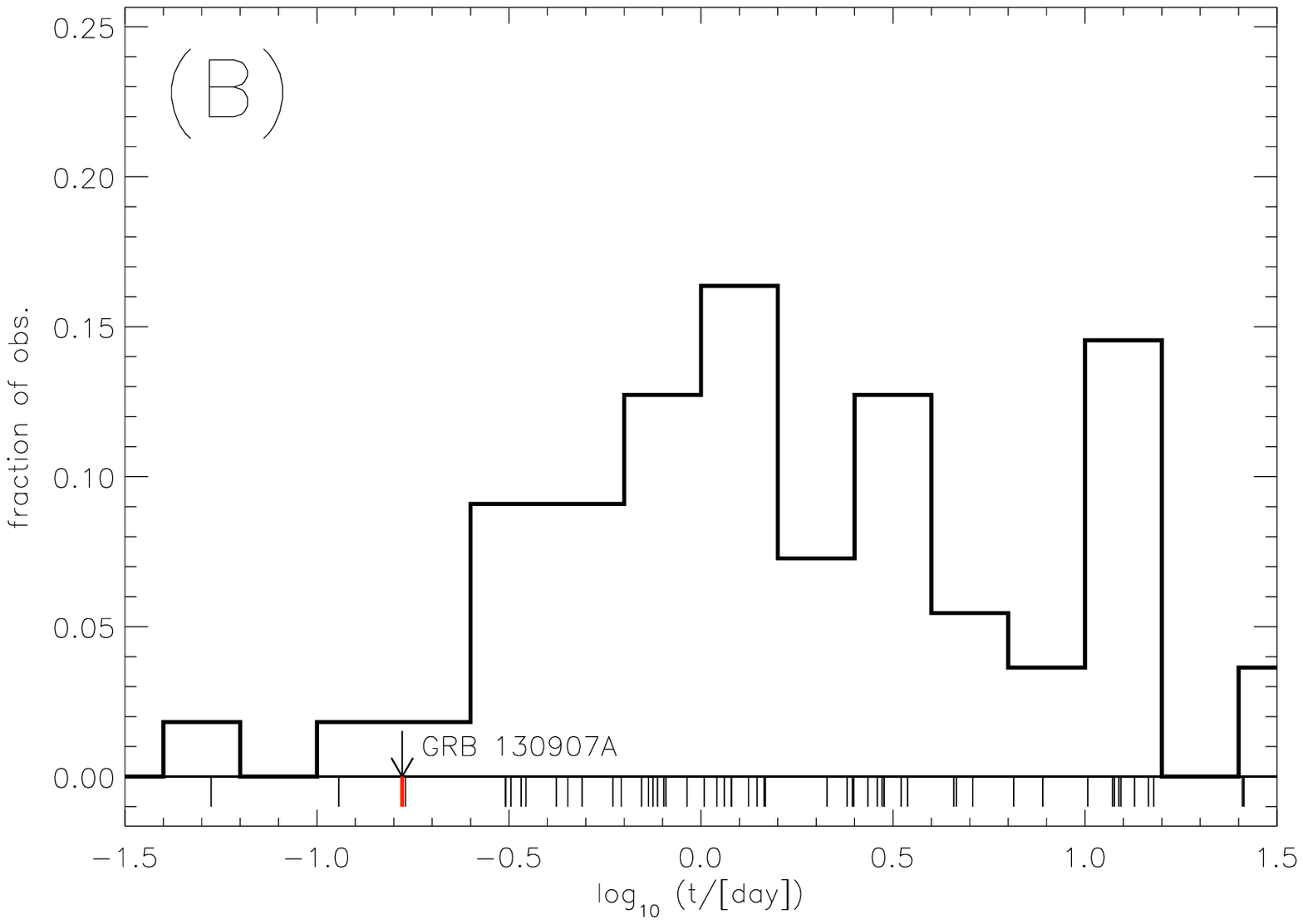}
\caption{{\tiny(A) Distribution of time delays from $\gamma$-ray trigger to first
radio observation of GRBs from GRB\,970111 to GRB\,141109A. The histogram
represents the earliest \textit{observation} for a given burst. The plot shows
detections ($>3 \sigma$) out of 350 observations. The bulk of the observations
were carried out with the VLA, later Karl G. Jansky VLA. Other observatories
include: ATCA, Ryle telescope, AMI, WSRT, JCMT, MAMBO, OVRO and CARMA.  The rug
shows time delays for the \textit{detections} and the arrow indicates the time
delay to the first radio observation/detection of GRB\,130907A.  (B) Histogram
of first \textit{detections} (above $3 \sigma$) at frequencies greater than
$10\GHz$.  The total number of bursts detected at frequencies above 10\,GHz is
55. Delay times for bursts before GRB\,110731A are compiled in
\citet{Chandra2012}. For the others we have used:
GRB 141109A: \citet{2014GCN..17070...1},
GRB 141026A: \citet{2014GCN..17019...1},
GRB 140903A: \citet{2014GCN..16777...1},
GRB 140713A: \citet{2014GCN..16593...1},
GRB 140709A: \citet{2014GCN..16595...1},
GRB 140703A: \citet{2014GCN..16516...1},
GRB 140515A: \citet{2014GCN..16283...1},
GRB 140419A: \citet{2014GCN..16122...1},
GRB 140311A: \citet{2014GCN..15961...1}, and \citet{2014GCN..15985...1},
GRB 140304A: \citet{2014GCN..15931...1},
GRB 131224A: \citet{2013GCN..15626...1},
GRB 131108A: \citet{2013GCN..15478...1},
GRB 130912A: \citet{2013GCN..15227...1},
GRB 130907A: \citet{2013GCN..15200},
GRB 130822A: \citet{2013GCN..15122...1},
GRB 130702A: \citet{2013GCN..14990...1}, and \citet{2013GCN..14979...1},
GRB 130609A: \citet{2013GCN..14863...1},
GRB 130606A: \citet{2013GCN..14817...1},
GRB 130603B: \citet{2013GCN..14751...1},
GRB 130518A: \citet{2013GCN..14689...1},
GRB 130427A: \citet{2013GCN..14480...1}, and \citet{2013GCN..14494...1},
GRB 130418A: \citet{2013GCN..14387...1},
GRB 130215A: \citet{2013GCN..14210...1},
GRB 130131A: \citet{2013GCN..14172...1},
GRB 121226A: \citet{2012GCN..14126...1},
GRB 121024A: \citet{2012GCN..13903...1},
GRB 120923A: \citet{2012GCN..13813...1},
GRB 120804A: \citet{2012GCN..13587...1},
GRB 120802A: \citet{2012GCN..13578...1},
GRB 120729A: \citet{2012GCN..13547...1},
GRB 120521C: \citet{2012GCN..13336...1},
GRB 120404A: \citet{2012GCN..13231...1},
GRB 120327A: \citet{2012GCN..13180...1},
GRB 120326A: \citet{2012GCN..13175...1},
GRB 120305A: \citet{2012GCN..13010...1},
GRB 120119A: \citet{2012GCN..12895...1},
GRB 111215A: \citet{2011GCN..12710...1},
GRB 111209A: \citet{2011GCN..12804...1},
GRB 111117A: \citet{2011GCN..12571...1},
GRB 111022B: \citet{2011GCN..12496...1},
GRB 111008A: \citet{2011GCN..12436...1},
GRB 111005A: \citet{2011GCN..12422...1}.}
}
\label{fig:dthist} 
\end{center}
\end{figure}

Our observations of GRB\,130907A rank among the earliest VLA detections of a
GRB (see Figure \ref{fig:dthist}).  The follow-up started at {  $3.64$\, hours}
after the trigger.  Hereafter, we refer to the first VLA observation as Epoch I
or EI; later VLA observations are labeled incrementally up to Epoch V at 42\,d
after the trigger (Table\,\ref{tab:obs}). 

 VLA data were reduced and imaged
using the Common Astronomy Software Applications (CASA) package.  Specifically,
the calibration was performed using the VLA calibration pipeline v1.2.0. After
running the pipeline, we inspected the data (calibrators and target source) and
applied further flagging when needed.  3C286 and J1423+4802 were used as flux
and phase calibrators, respectively. The VLA measurement errors are a
combination of the rms map error, which measures the contribution of small
unresolved fluctuations in the background emission and random map fluctuations
due to receiver noise, and a basic fractional error \citep[here conservatively
estimated to be $\approx 5\%$, based on the flux variations measured for
the phase calibrator;][]{Ofek2011} which accounts for inaccuracies of the
flux density calibration. Theses errors were added in quadrature and total
errors are reported in Table 2. { An additional source of error in the radio
band can occur from scintillation, which we discuss in Section \ref{sec:scint}.
}

{ 
We also observed the position of GRB\,130907A with the Combined Array for Research
in Millimeter Astronomy (CARMA) on 2013-09-08 between 21:46:36 and 23:12:35 UT
($t_{\rm mid} = 1.03$ day).  Observations were conducted in single-polarization
mode with the 3 mm receivers tuned to a frequency of 93 GHz, and were reduced
using the Multichannel Image Reconstruction, Image Analysis and Display
environment (MIRIAD).  Flux calibration was established by a short observation
of Mars at the beginning of the track.  We detect no source at the location of
the GRB afterglow in the reduced image, with a limiting flux of 1.1 mJy
(2$\sigma$).}

\begin{figure}[htbp]
\begin{center}
\includegraphics[width=0.999\columnwidth,angle=0]{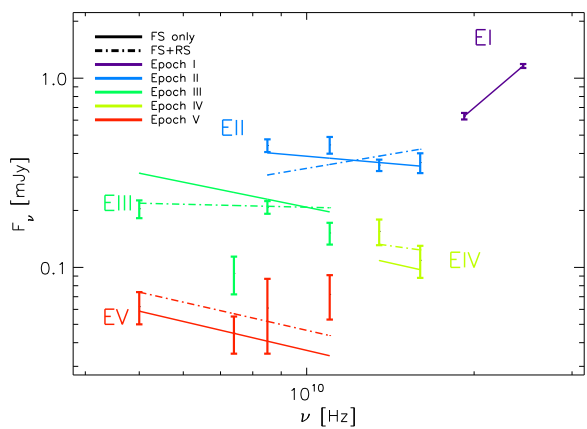}
\caption{A compilation of VLA measurements for GRB\,130907A,  displaying the
spectral evolution and model fits based on the forward shock only, and forward
+ reverse shock (see Figure \ref{fig:radiolc}). EI through EV mark the five
epochs (see Table\,\ref{tab:obs}). 
} 
\label{fig:radiospec} 
\end{center}
\end{figure}

\begin{figure*}[htbp]
\begin{center}
\includegraphics[angle=0]{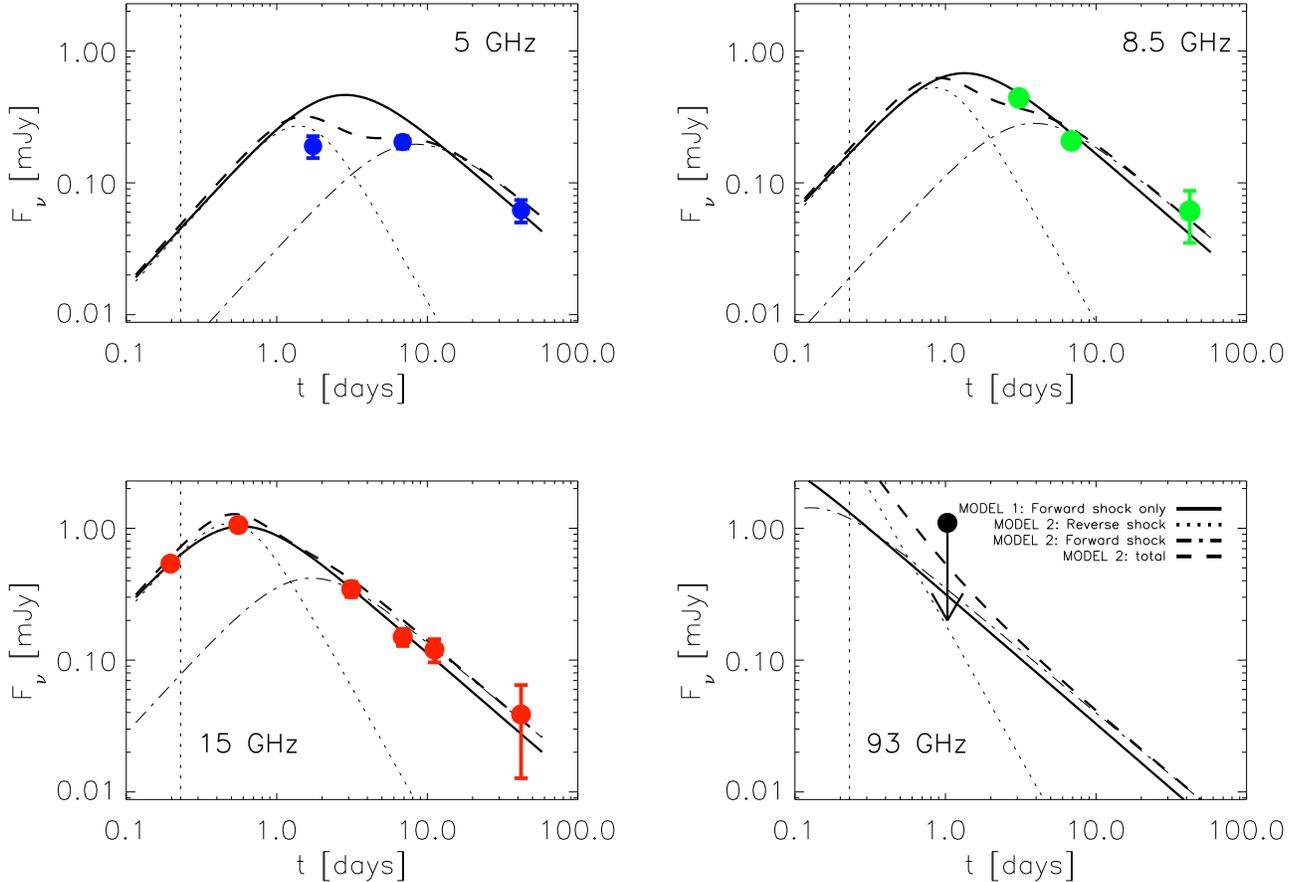}
\caption{Radio light curves of GRB\,130907A at 5, 8, 15 and 93 GHz.  Continuous
lines show the model consisting only of the forward shock.  Dotted and
dot-dashed lines represent models when both the reverse shock and the forward
shock are present (their sum is plotted with dashed lines). The forward shock
rising slope is $\alpha=-5/4$, while the reverse shock rises with
$\alpha=-9/7$.  In the  ISM case, the decay slope for the forward and reverse
shock is $\alpha\approx0.87=3(p-1)/4$ and $\alpha=(27p+7)/35\approx2.04$
respectively.  The vertical dotted line marks the transition from wind to ISM.
In the wind case, the slopes of the rising part of the light curve are steeper
than in the ISM case: $\alpha=-7/4$ (forward shock) and $\alpha=-65/42$
(reverse shock). Because there is only one data point in the wind regime we
don't plot these cases separately.  We obtain 5, 8.5 and 93 GHz light curves by
scaling the 15 GHz light curve using the well-known synchrotron radiation
scalings.  }
\label{fig:radiolc} 
\end{center}
\end{figure*}

During Epoch I, the radio spectral index is $\beta^I_{\rm radio}=-2.50\pm0.19$,
which strongly suggests that Epoch I is self-absorbed. As evident from Figure
\ref{fig:radiospec}, at later times our VLA observations suggest an evolution
of the radio emission toward an optically thin regime, with the spectral index
progressively becoming flatter with time.  In order to extract from our data a
well sampled radio light curve, we have extrapolated VLA measurements at
various epochs to 15\,GHz { using the best fitting power law to the spectra
at a given epoch} (Table\,\ref{tab:obs}). For fitting purposes, we set the
initial light curve slope to $\alpha=-7/4$, corresponding to the temporal
behavior expected for a fireball expanding in a wind environment (see
Section\,\ref{sec:model}). With this choice, we find $\alpha=0.87\pm 0.07$ for
the late-time temporal slope at 15\,GHz, and a peak-time of $t_{\rm
r,pk}=0.38\pm 0.03$ days (Figure\,\ref{fig:radiolc}).

\begin{longtable}{cccccc} \hline \hline 
Time$_{\rm mid}$[days] & Filter & mag[AB]  & Instrument & Reference \\
 \hline 
$5\times10^{-4}$ 	 & r  & 15  $\pm$ (0.2) &  MASTER & (1) \\
0.0152	& Ic 	& 15.26 $\pm$ 0.03 & Tautenburg& (2) \\
0.0172	& Rc 	& 16.57 $\pm$ 0.03 & Tautenburg& (2) \\
0.127	& I 	& 18.27 $\pm$ 0.15&  Skynet& (3)\\
0.131	& I 	& 18.76 $\pm$ 0.23&  Skynet& (3) \\
0.132	& I 	& 18.41 $\pm$ 0.084&  Skynet& (4) \\
0.136	& I 	& 18.40 $\pm$ 0.12&  Skynet& (3) \\
0.144	& R 	& 19.65 $\pm$ 0.20&  Skynet& (3) \\
0.148	& R 	& 19.51 $\pm$ 0.13&  Skynet& (4) \\
0.154	& i' 	& 18.86 $\pm$ 0.09&  Skynet& (4) \\
0.158	& I 	& 18.35 $\pm$ 0.13&  Skynet& (3) \\
0.159	& r' 	& 19.65 $\pm$ 0.10&  Skynet& (4) \\
0.164	& I 	& 18.91 $\pm$ 0.16&  Skynet& (3) \\
0.167	& I 	& 18.79 $\pm$ 0.09 &  Skynet& (4) \\
0.217	& R 	& 20.06 $\pm$ 0.47 &  T21& (5) \\
0.232	& r 	& 20.01 $\pm$ 0.03 &  RATIR& (6) \\
0.232	& i 	& 19.30 $\pm$ 0.02 &  RATIR& (6) \\
0.258 	& Ic 	& 19.05 $\pm$ 0.09&  Skynet& (4) \\
0.287	& Rc 	& 19.91 $\pm$ 0.11&  Skynet& (4) \\
0.316	& Ic 	& 19.31 $\pm$ 0.13&  Skynet& (4) \\
0.341	& Ic 	& 19.51 $\pm$ 0.22&  Skynet& (4) \\
2.22	& r 	& 21.92 $\pm$ 0.12 &  RATIR& (7) \\
2.22	& i 	& 21.38 $\pm$ 0.09 &  RATIR& (7) \\
2.56	& R 	& 21.44 $\pm$ 0.16 &  Maidanak& (8) \\
3.22	& r 	& 22.40 $\pm$ 0.14 &  RATIR& (9) \\
3.22	& i 	& 21.73 $\pm$ 0.10 &  RATIR& (9) \\
\hline
8.56$\times10^{-4}$ 	 & white 	 &  15.45 $\pm$ 0.02  & UVOT &  (10) \\ 
7.13$\times10^{-3}$	 & v 	 &  16.29 $\pm$ 0.16  & UVOT &  (10) \\
6.27$\times10^{-3}$	 & b 	 &  16.78 $\pm$ 0.12  & UVOT & (10) \\
3.31$\times10^{-3}$	 & u 	 &  15.87 $\pm$ 0.04  & UVOT &  (10) \\
7.71$\times10^{-3}$	 & w1 	 &  18.54 $\pm$ 0.31  & UVOT & (10) \\
7.42$\times10^{-3}$	 & m2 	 &  $>$19.20  		& UVOT & (10) \\
6.85$\times10^{-3}$	 & w2 	 &  $>$18.70	 		& UVOT & (10) \\

0.232 	 & Z 	 &  18.78 $\pm$ 0.05  & RATIR & (6) \\ 
0.232 	 & Y 	 &  18.48 $\pm$ 0.06  & RATIR & (6) \\
0.232 	 & J 	 &  18.13 $\pm$ 0.06  & RATIR & (6) \\
0.232 	 & H 	 &  17.63 $\pm$ 0.05  & RATIR & (6) \\ 

2.22 	 & Z 	 &  $>$21.37  & RATIR &  (7)\\ 
2.22 	 & Y 	 &  $>$20.87 	& RATIR & (7)\\ 
2.22 	 & J 	 &  $>$20.58  & RATIR & (7)\\ 
2.22 	 & H 	 &  $>$19.97  & RATIR & (7)\\ 
3.22 	 & Z 	 &  $>$21.80  & RATIR & (7)\\ 
3.22 	 & Y 	 &  $>$21.05  & RATIR & (7)\\ 

5.23 	 & r 	 &  $>$22.68   & RATIR &  (11) \\ 
5.23 	 & i 	 &  $>$22.62   & RATIR &  (11) \\ 
5.23 	 & Z 	 &  $>$21.30   & RATIR &  (11) \\ 
5.23 	 & Y 	 &  $>$20.70   & RATIR &  (11) \\ 
5.23 	 & J 	 &  $>$20.42   & RATIR &  (11) \\ 
5.23 	 & H 	 &  $>$19.69   & RATIR &   (11)\\ 
\hline
\caption{Optical observations of GRB\,130907A. (References: 1: \citet{2013GCN..15184}, 
2: \citet{2013GCN..15194},
3: \citet{2013GCN..15191},
4: \citet{2013GCN..15193},
5: \citet{2013GCN..15201},
6: \citet{2013GCN..15192},
7: \citet{2013GCN..15208},
8: \citet{2013GCN..15240},
9: \citet{2013GCN..15209},
10: \citet{2013GCN..15195},
11: \citet{2013GCN..15223}.
)}

\label{tab:optical}
\end{longtable}

\begin{table}
\begin{center}
\begin{tabular}{ccccc} \hline \hline 
Time$_{\rm mid}$ [days] & $\nu$ [GHz] & Flux [$\mu$Jy] & Instrument&Reference\\
 \hline 
0.193 (EI)	&  19.2 	&  630 $\pm$ 25 	& EVLA 	& (this work) \\
0.193  (EI) 	&  24.5 	&  1160 $\pm$ 28 	& EVLA 	& (this work) \\
0.550  	  	&  15 		&  1060 $\pm$ 110 	& AMI 	& (1) \\
1.03		&	93		&  $<$1100 (2$\sigma$) & CARMA & (this work)\\
1.735 	  	&  5 		&  190 $\pm$ 30 	& WSRT 	& (2) \\
3.08  (EII) 	&  8.5 		&  441 $\pm$ 34 	& EVLA 	& (this work) \\
3.08 (EII)	&  11 		&  444 $\pm$ 45 	& EVLA 	& (this work) \\
3.08 (EII)	&  13.5 	&  347 $\pm$ 24 	& EVLA 	& (this work) \\
3.08 (EII)	&  16 		&  358 $\pm$ 43 	& EVLA 	& (this work) \\
6.86 (EIII)	&  5 		&  204 $\pm$ 22 	& EVLA	& (this work) \\
6.86 (EIII)	&  7.4 		&  93 $\pm$ 21 		& EVLA	& (this work) \\
6.86 (EIII)	&  8.5 		&  208 $\pm$ 16 	& EVLA	& (this work) \\
6.86 (EIII)	&  11 		&  152 $\pm$ 20 	& EVLA	& (this work) \\
11.15 (EIV) 	&  13.5 	&  155 $\pm$ 24 	& EVLA	& (this work) \\
11.15 (EIV) 	&  16 		&  109 $\pm$ 21 	& EVLA	& (this work) \\
41.96 (EV)	&  5 		&  62 $\pm$ 12 		& EVLA	& (this work) \\
41.96 (EV)	&  7.4 		&  45 $\pm$ 10 		& EVLA	& (this work) \\
41.96 (EV)	&  8.5 		&  61 $\pm$ 26 		& EVLA	& (this work) \\
41.96 (EV)	&  11 		&  72 $\pm$ 19	 	& EVLA 	& (this work) \\
\hline
\end{tabular}
\caption{Radio observations of GRB\,130907A. Times are calculated since the
$\gamma$-ray trigger.  See Figure \ref{fig:radiospec} for a plot of the VLA
observations. (EI)-(EV) indicate the five observing epochs with the VLA.
References: (1) \citet{2013GCN..15211}, (2) \citet{2013GCN..15207} }

\label{tab:obs}
\end{center}
\end{table}

\subsection{Radio-to-GeV spectral energy distribution}
\label{sec:radio2GeV}
\begin{figure}[htbp]
\begin{center}
\includegraphics[width=0.999\columnwidth,angle=0]{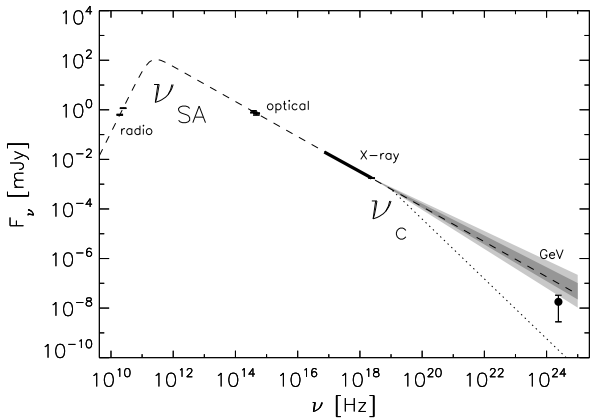}
\caption{ Spectral energy distribution of GRB\,130907A during Epoch I (0.19
days after the trigger; see Table \ref{tab:obs}) from radio to 10 GeV.  The
broken power-law spectral indices are $\beta=\{-2.5, 0.69\}$ for the dashed
line. $\beta=1.19$ for the dotted line. $\nu_{\rm SA}$ is the self-absorption
frequency and $\nu_c$ is the cooling frequency.  The dark and light shaded
regions mark the 1$\sigma$ and 2$\sigma$ uncertainties, respectively, of the
extrapolated X-ray spectrum.} 
\label{fig:e1} 
\end{center}
\end{figure}

During Epoch I, defined by the time of the first VLA observation, we have a
spectral coverage spanning $\sim 15$ orders of magnitude from radio to
$\gamma$-rays (Figure\,\ref{fig:e1}).  The spectral index in the radio band is
$\beta^I_{\rm radio}=2.50\pm0.19$, which indicates that the radio emission is
self-absorbed at this epoch (Section\,\ref{sec:radiolc}).  We extrapolate the
optical measurements to Epoch I using the temporal indices as derived in
Section\,\ref{sec:optdata}, i.e. $\alpha_{\rm o}^{I}\approx1.17-1.37$.  The
break in the spectrum at the intersection of the extrapolated radio and optical
measurements, occurs at $\nu_{SA} \approx 2\times 10^{11} \Hz$.

In the X-rays, the spectral and temporal slopes during Epoch I are
$\beta^I_{\rm X}=0.69\pm0.05$ and $\alpha_{\rm X}^{I}=1.32\pm 0.05$,
respectively (see Section\,\ref{sec:xlc}).  The emission in the LAT energy band
has a temporal index of $\alpha^I_{\rm LAT}=1.13\pm 0.57$ \citep{Tang+14ic},
which is consistent with the X-ray one within the large uncertainties. 

Interestingly, the spectrum at Epoch I is consistent with a single power-law
component (dashed line in Figure \ref{fig:e1}) from optical to the GeV range.
It should be noted, however, that the low photon counts do not allow to derive
a spectral index for the GeV emission, thus a spectral break might be present
between the X-rays and the GeV range (dotted line in Figure \ref{fig:e1}).  In
other words, the consistency of the X-ray and GeV (LAT) temporal indices does
not require the presence of a spectral break, but the uncertainties in the LAT
flux do not exclude the presence of a cooling break ($\nu_c$) between the
X-rays and GeV band. { In fact, as we discuss in Section \ref{sec:micro}, a
cooling break just above the X-ray band during Epoch I (dotted line) helps
explain the late-time spectral evolution observed in X-rays. The presence of
such a break would cause the LAT flux to be slightly underpredicted (by $\sim
1.2 \sigma$), but this could be mitigated by invoking an emergent SSC component
\citep[as suggested by][]{Tang+14ic}.}

\section{Modeling}
\label{sec:model}
\subsection{Initial considerations}
\label{sec:initial}
We assume the radiation originates from synchrotron radiation of shock
accelerated electrons.  The electrons have a distribution described by a broken
power-law.  The resulting synchrotron spectrum is also a set of joined
power-laws with breaks at the characteristic frequencies: the injection
frequency, $\nu_m$, where the bulk of the electrons radiate; the cooling
frequency, $\nu_c$, where the cooling time of the electron radiating at this
frequency is equal to the dynamical time; and the self-absorption frequency,
$\nu_{\rm SA}$, which is defined as the frequency where the optical depth for
synchrotron photons becomes greater than unity for scattering on the
synchrotron emitting electrons \citep{Meszaros+97ag,Sari+98spectra,
Granot+02Dabreaks}.

First, inspecting the general properties of the multi-wavelength afterglow
observations, this burst presents a conundrum. As we explain in what follows,
based on the closure relations between the temporal and spectral indices for
GRBs \citep[see e.g.  ][]{Racusin+11fermiswift}, the early ($t \lesssim 0.2
\,{\rm d}$) X-ray light curve slope is suggestive of a wind environment and
consistent with the optical measurements.  On the other hand, the radio
observations are better explained in an ISM (Section \ref{sec:csm}).  Finally,
the late-time X-ray light curve has a steep slope (Figure\,\ref{fig:Xlc}) and a
strong spectral evolution (Figure\,\ref{fig:spidx}) indicating the passage of a
characteristic frequency through the X-ray band (although the steep slope of
the X-ray light curve is difficult to explain in both an ISM and a wind
environment).

We thus suggest that a simple external-shock model is not able to account for
all the observed data.  Among the multitude of extensions to the simplest
model, a possible explanation is that GRB\,130907A is produced by a shock
initially propagating in a wind environment, which then transitions to a
constant density ISM.  Similar models were proposed by e.g.
\citet{Wijers01transition, Peer+06rs,Gendre+07transition, Kamble+07transition}.
Hereafter we assume that the spectral evolution observed in the late-time X-ray
afterglow is due to the passage of a characteristic frequency in band. However,
we also note that the higher-than-average reddening observed in GRB\,130907A
suggests this burst might be a good candidate for the dust screen scenario
proposed by \citet{Evans+14dust} to explain the spectral evolution observed in
X-rays for GRB\,130925A.

As the blast wave transitions from a wind to an ISM environment, roughly at the
time of the X-ray break, the cooling frequency ($\nu_c$) sweeps through the
X-rays causing the observed spectral evolution. The passage of the cooling
frequency will not affect the radio light curve which behaves simply as in the
case of an ISM. 

By looking at the 15\,GHz light curve (Figure \ref{fig:radiolc}), we find no
obvious requirement to include a reverse shock in our modeling. However, the
behavior of the 5\,GHz flux seems to favor a forward-plus-reverse-shock model.
A better temporal coverage, particularly between 1\,d and 2\,d since trigger at
the highest radio frequencies, would likely have allowed us to securely
discriminate between a forward-shock-only and a forward-plus-reverse-shock
scenario.

\subsection{Early-time wind profile}
\label{sec:csm}
If we assume that, before the temporal break ($t < 0.23\,{\rm d}$), the X-ray
band is below the cooling frequency and above the self-absorption and injection
frequencies ($\nu_m , \nu_{\rm SA}<\nu_X<\nu_c$), based on the spectral and
temporal indices, we can estimate the nature of the interstellar matter density
profile ($\rho\propto R^{-k}$) e.g. from \citet{Sari+00refresh}:
$k=4/[1+1/(2\alpha-3\beta)=1.48\pm0.30$, which is suggestive of a wind
environment before $t_{\rm break}$\footnote{For a detailed treatment of the
radiation from GRB afterglows in a general circumstellar density profile, see
\citet{Yi+13stratified}.}. This results in a power-law index of the electron
distribution of $p\approx2.37\pm0.10$ ($p=2\beta+1$). 

The temporal evolution of the cooling frequency is $\nu_c\propto t^{-\alpha}$,
where $\alpha={-(4-3k)/(8-2k)}\approx-0.09$, for a circumstellar density
profile index $k\approx1.5$. Thus, the cooling break is almost constant with
time,  similarly to what was found by \citet{Perley2014} in the case of
GRB\,130427A.

The self-absorbed spectrum at Epoch I is a noteworthy feature of this burst,
and the spectral index is $-2.5$ is unique: The more commonly discussed cases
for synchrotron emission have a self-absorbed slope of $2$
\citep[e.g.][]{Yost+02sa}. The evolution of the self-absorption frequency
provides another argument in favor of the wind nature of the environment closer
to the explosion site.  At Epoch I, $\nu_{\rm SA}^{I}(0.19\,{\rm
d})\approx200\,$GHz (though this value should be considered uncertain due to
the extrapolation over many orders of magnitude; see  Figure\,\ref{fig:e1} and
Section\,\ref{sec:radio2GeV}).  Epoch II is clearly not self-absorbed
(Figure\,\ref{fig:radiospec}), thus  $\nu_{\rm SA}^{II}(3.1\,{\rm d})\lesssim
10\,$GHz. In an ISM environment, $\nu_{SA}\propto t^{-(3p+2)/(2(p+4))}$, which
yields a $\nu_{\rm SA}^{II}(3.1\,{\rm d})\gtrsim 20$\,GHz for $p$ between 2 and
3.  On the other hand, for a wind environment, $\nu_{SA}\propto
t^{-(3(p+2))/(2(p+4))}$, which yields a self-absorption frequency below 10 GHz
at Epoch II, in agreement with our observations.

We finally note that a more common self-absorbed spectral slope of $-2$, which
would be expected in the $\nu_{\rm radio}<\nu_{\rm SA}~(<\nu_m<\nu_{\rm
optical})$ regime, would be consistent with the observed value of $\beta^I_{\rm
radio}=-2.50\pm0.19$ only at the $\approx 2.6\sigma$ level.  In this regime,
because the later epochs are not self-absorbed, $\nu_{\rm SA}$ would need to
pass in the radio band by the time of Epoch II. In the ISM case, $\nu_{\rm
SA}\propto t^0$ while in the wind case $\nu_{\rm SA}\propto t^{-3/5}$. Thus,
our conclusion favoring an initial wind environment is not affected by the
relative ordering of $\nu_{\rm SA}$ and $\nu_m$. 

\subsection{Spectral energy distribution at Epoch I}
The spectrum at Epoch I suggests a synchrotron origin for the entire spectral
range \citep[see e.g.  ][]{Kumar+09gevforsho, Kouv+13nustar}. However, in a
synchrotron model one has to overcome the maximum attainable synchrotron energy
condition, which might pose a problem for a synchrotron-only model
\citep{dejager+96maxsyn}.  A scenario for interpreting Epoch I spectrum with a
forward-shock synchrotron component is in the regime where: $\nu_m<\nu_{\rm
radio}<\nu_{\rm SA}<\nu_{\rm opt}< \nu_X <\nu_c< \nu_{\rm LAT}$ (dotted line in
Figure \ref{fig:e1}).  Although observations are equally consistent with
$\nu_{\rm LAT}<\nu_c$ (dashed line in Figure \ref{fig:e1}), the spectral
evolution observed in the late-time X-ray light curve favors a model where
$\nu_c$ lies just above the \textit{Swift}/XRT range during Epoch I
(Section\,\ref{sec:xbrk}). For the implications of an alternative
inverse-Compton model, see: \citet{Tang+14ic}.

\subsection{ISM transition and origin of the X-ray break}
\label{sec:xbrk}
The break observed in X-rays (Figure\,\ref{fig:Xlc}) is clearly inconsistent
with a jet break because it is chromatic: It only occurs in X-rays, the radio
component does not have a break, while the optical measurements do not support
it. Chromatic breaks are not uncommon in GRB afterglows \citep[see e.g.
][]{panaitescu06b,Liang+07shallow}, but are not generally accompanied by
spectral evolution. 

No clear achromatic break can be identified in the X-ray light curve until
about 20 d since explosion. By imposing $t_{\rm jet}>20$ d, we get $\theta_j
\gtrsim 12^\circ E_{54}^{-1/8} (n/50 \cm^{-3}) ^{1/8} (t/20\,d)^{3/8}
((1+z)/2.23)^{-3/8} $ and a beaming-corrected energy of $E_{\rm jet}\gtrsim
3.4\times 10^{52} \erg$.

To account for the observed spectral evolution, we assume the cooling frequency
starts to cross the \textit{Swift}/XRT band at about the same time the shock
reaches the transition from the wind environment to the ISM.  This requires
that  the cooling break lies just above the XRT range during Epoch I, since in
a $k=1.5$ medium $\nu_c$ is almost constant with time (Section\,\ref{sec:csm}),
while $\nu_c \propto t^{-1/2}$ in an ISM.  Indeed, the \textit{Swift}/XRT data
imply that $\nu_c$ takes about two orders of { magnitude} in time (from
$2\times10^4 \s$ to $2\times 10^6 \s$) to move through approximately one order
of magnitude in frequency (from 10 keV to 0.3 keV, the range of XRT; see
Figure\,\ref{fig:spidx}).  Moreover, the observed change in spectral slope,
$\Delta\beta=0.9\pm 0.4$, is consistent with the theoretical expectations for
the passage of $\nu_c$ in band, $\Delta \beta=0.5$ (Figure \ref{fig:spidx}).
{ We note that a constant density medium might not necessarily be related to
ISM, but it might also result from the interaction of the stellar wind with the
circumstellar environment. The latter may indeed be homogenized by this
interaction, as suggested by several studies
\citep[e.g.][]{vanMarle+06termrad}}.

Last but not least, a transition from a wind-like ($k\approx1.5$) medium to an ISM
also allows us to explain the observed late-time slope of the radio light curve.
Indeed, for a $k=1.5$ medium to yield a radio temporal slope of $\alpha\sim
0.9$ (Section\,\ref{sec:radiolc}), one needs an electron distribution index of
$p\approx 1.8$\footnote{In order to obtain a finite energy in electrons it is
usual to assume $p>2$. See however \citet{Panaitescu01harde} for a treatment of
a $p<2$ case.}. On the other hand, in an ISM, the observed late-time radio
light curve slope implies $p=4/3\alpha+1=2.2\pm0.1$, which is consistent with
the value independently derived from the early-time X-ray observations
(Section\,\ref{sec:csm}).

A shortcoming of our presented model is that the X-ray temporal slope after the
break is steeper than expected for a transition of the cooling frequency.  In
the framework of a simple synchrotron model the steepest temporal decay index
is $(3p-2)/4$ and it does not depend on the nature of the environment. This
expression for a temporal index applied to the X-ray data yields $p\approx4.1$,
which is an unusually steep value for the electron distribution index.

With the introduction of a narrow jet responsible for the X-ray emission and a
wider jet for the longer wavelength afterglow \citep{racusin08,
vanderhorst+14doublejet}, one would decouple the X-ray and optical/radio
behavior.  This way a break in the X-ray would be due to the narrow jet with
opening angle $\theta_j\sim 2^\circ E_{54}^{-1/4} A_{*,-1}^{-1/4}
(t/0.23\,d)^{1/4} ((1+z)/2.23)^{-1/4}$, and energy release corrected for
beaming of $E_{\rm jet}\approx 10^{51} (\theta/2^\circ)^2 \erg$.  { This is
the opening angle and the beaming corrected energy respectively if we interpret
the break in X-ray as a jet break. } The consequent passing of the cooling
frequency would explain the spectral evolution. The radio and optical fluxes
would be due to synchrotron radiation from a wider jet.  This would naturally
explain the late $t^{-2.57}$ behavior of the X-ray light curve with $p\approx
2.57$, as the post jet-break phase flux evolves as $t^{-p}$. We consider this
model lacks solid observational evidence, but can be substantiated for similar
future bursts with better temporal coverage of the emission.

There are other possible solutions to the puzzling late-time behavior of
GRB\,130907A X-ray light curve (see Section \ref{sec:xbrk}): the break in the
light curve can be attributed to the end of the shallow decay (plateau) phase
\citep{Nousek:2006ag,Liang+07shallow} of the GRB. It is uncommon, but not
impossible that the plateau phase has a temporal index of $1.3$ \citep[see
Figure 4 in][]{Grupe+13plateau}. In this case the flux before the break is
related to central engine activity, and the evidence for an initial wind
environment is not as compelling. The discrepancy between the late radio and
X-ray temporal indices is still too large to be solved by this model. 

A rather contrived setup would ascribe the radio emission to the usual
synchrotron component propagating in the ISM and the X-ray flux would be given
by synchrotron self-Compton radiation. The self-Compton light curves can be as
steep as $t^{(9p-11)/8}$ for ISM to  $t^{-p}$ in wind case. It is not possible
to constrain a self-Compton component form the observations, but one would need
to imagine a fine tuned interplay between the synchrotron and SSC components to
explain the observed X-ray slopes at late times.

\subsection{Transverse size of the jet and scintillation}
\label{sec:scint}
Within the limits of our temporal coverage of the radio afterglow of GRB
130907A, no strong flux modulation due to radio scintillation is apparent. In
this section, we verify a posteriori that the size of the expanding shell
derived from our modeling, which assumes negligible scintillation effects, is
indeed consistent with this assumption.

In a wind environment, the angle subtended by the fireball can be calculated
as: $\theta_S=2R_\perp/D_A\approx 5.9 \mu{\rm as} (E_{{\rm
iso},54} / A_{*,-1})^{1/4} (t/t_I)^{3/4} $, where $E_{\rm iso}$ is the isotropic
equivalent { kinetic} energy of the GRB, $A_*$ is a parameter describing the
ratio of stellar mass loss to the wind velocity ($A_*= (\dot{M}/10^{-5} M_\odot
{\rm yr}^{-1} v_8^{-1}$), $2 R_\perp$ is the transverse size of the jet,
$D_A=D_L/(1+z)^2$ is the angular diameter distance.

We use the above estimate { for the size of the expanding shell} to evaluate
the effects of scintillation on our VLA observations \citep{Walker98scint,
Walker01scint}.  At the position of GRB\,130907A (galactic coordinates
$(l,b)\approx(62^\circ,55^\circ)$), the upper-limit frequency for the strong
scattering regime is $\nu_0\approx 7\GHz$.  At this frequency the source is
significantly affected by scintillation if its size is smaller than $\theta_0
\approx 4 \mu{\rm as}$. Scintillation affects the lower frequency observations
more. Also, the projected size of the fireball is expected to increase with
time, so later observations are less affected. 

Scintillation introduces a random scatter around the 'real' value of the flux.
To characterize the strength of this scattering, it is customary to provide the
root mean square of the fluctuations for a given frequency and a given size of
the emitter.  The root mean square of the scattering is given by:
$m=m_p(\theta_S/\theta_F)^{-7/6}=0.08$ where $m_p= (\nu/\nu_0)^{-17/12}$ and
$\theta_F=\theta_0 (\nu/\nu_0)^{-1/2}$. Here $\nu$ is the lowest frequency in
the earliest radio observation ($\nu=19.2$\,GHz) which is the most affected by
scintillation. { The low value of $m$ suggests the Epoch I observations are not
affected significantly by scintillation.} The AMI observations were also
carried out at an early time, thus could possibly be affected by scintillation.
At $0.55$\,d after the trigger, at $15 \GHz$, similar calculations yield
$m\approx0.06$.  The $5\GHz$ WSRT observation at $1.74$\,d falls into the
strong scattering regime. The theoretically expected rms caused by
scintillation will be $m(5 \GHz)=(\nu/\nu_0)^{17/30}
(\theta_s/\theta)^{-7/6}\approx0.2$. This measurement may thus be the  most
affected by scintillation, so we add an uncertainty of $20\%$ in quadrature to
its error. Based on our model for GRB\,130907A, all other bands are expected to
have insignificant distortions by scintillation.

\subsection{Emission radius from self-absorption}
The self-absorption frequency occurs where the synchrotron emitting electrons
have the same energy that corresponds to the comoving brightness temperature
($T'$) of their emitted radiation. We have $\nu_{\rm SA}>\nu_m$ and
$kT'=\gamma_{\rm SA}m_e c^2$.  Using this, we can constrain the radius of
emission from \citep[see e.g.][]{Shen+09SA}: $F_{\nu_{\rm SA}} (D_L/R)^2 =
2\gamma_{\rm SA} m_e c^2 \Gamma/(1+z)^3$ and get: $R=6.5\times 10^{17}\cm~
A_{*,-1}^{1/8}\epsilon_{B,-2}^{1/8}$.  In deriving this result we used: $
\gamma_{\rm SA}=(16 m_e c (1+z)\nu_{\rm SA}/3 q_e B \Gamma)^{1/2}$ for the
electron random Lorentz factor emitting at the $\nu_{\rm SA}$ frequency,
$B\approx 1 \G~ \epsilon^{1/4}_{B,-1} A_{*,-1}^{1/4} E_{54}^{1/4}
(t/t_I)^{-1/4}$ is the magnetic field strength and $\Gamma\approx 26~
A_{*,-1}^{-1/4} E_{54}^{1/4} (t/t_I)^{-1/4}$ is the bulk Lorentz factor,
consistent with the calculation of \citet{Anderson+14AMI}.

We note that the above calculated radius is a factor of $\Gamma(t)$ larger than
the transverse radius calculated for addressing the effects of scintillation
(Section \ref{sec:scint}). 

\subsection{Radius of the termination shock} 
As discussed above, we interpret the break in the X-ray light curve as the
transition between the wind and ISM environments.  The transition radius at the
time of the X-ray break ($t_{\rm X,break}\approx 2\times 10^4 \s$) from
\citet{panaitescu04} is: $R_t\approx 7.7\times 10^{17} \cm~ E_{54}^{1/2}
A_{*,-1}^{-1/2} (t/t_{\rm X,break})^{1/2}$. Keeping in mind the uncertainty on
the time of the X-ray break, this is broadly consistent with the radius derived
from self-absorption.  This radius is also close to values obtained by similar
calculations \citep{Kamble+07transition,Jin+09bubble}.

An approximate lower limit to the density of ISM
medium can be calculated  by substituting the transition radius in the
expression of the wind density. We get $n\gtrsim 1.2 \cm^{-3}$. More
accurately, from the expression of the cooling frequency (which we require to
be $\sim 10 \keV$ at $t=0.23$\,d) we have: $n\approx 50 \cm^{-3}
E_{54.2}^{-1/2}\epsilon_{B,-5}^{-3/2} (t/0.23\,d)^{-1/2}$ which is a reasonable
value for long GRBs. 

\subsection{Constraining the micro-physical parameters}
\label{sec:micro}
We assume the light curve at $15$\,GHz peaks due to the passage of the
self-absorption frequency through the $15\,$GHz band ($\nu_{\rm SA}$, see
Figure \ref{fig:radiolc}).

Here, using the expression of the peak frequency and flux at the time of the 
peak in the 15\,GHz light curve (which occurs in the wind environment), we get:
\begin{equation}
\nu_{\rm SA} =11\GHz~ E_{54}^{0.03}\epsilon_{e,-1}^{0.43} \epsilon_{B,-2}^{0.34} 
A_{*,-1}^{0.63} (t/t_{\rm r,pk})^{-1.03}\approx15 \GHz
\label{eq:nusa}
\end{equation}
and
\begin{equation}
F_{\nu_{\rm SA}}= 0.6~{\rm Jy}~ E_{54}^{0.82}\epsilon_{e,-1}^{1.08}
\epsilon_{B,-2}^{0.61} 
A_{*,-1}^{0.57} (t/t_{\rm r,pk})^{-0.82}\approx 1.1~ {\rm mJy}
\label{eq:fnusa}
\end{equation}
The exponents are for $p=2.38$ and have a strong dependence on the value of
$p$.

By assuming $\nu_m \lesssim 20 \ghz$ at the time of the first radio
observation, we get from the expression of $\nu_m$ for a wind medium
\citep[e.g.][]{Granot+02Dabreaks} 
\begin{equation}
\nu_m= 1.2\times 10^6 \GHz~ E_{54}^{1/2}
\epsilon_{e,-1}^2 \epsilon_{B,-2}^{1/2}\lesssim 20\GHz.
\label{eq:num}
\end{equation}

Roughly at the time of the temporal break in X-rays, which we associate with
the transition from wind to ISM, the cooling frequency starts to sweep through
the X-ray band of the XRT instrument ($0.3-10\keV$). Thus we will have: 
\begin{equation}
\nu_c=82 \eV E_{54}^{1/2} A_{*,-1}^{-2} \epsilon_{B,-2}^{-3/2} \sim 10 \keV.
\label{eq:nuc}
\end{equation}

By solving the above set of equations ((\ref{eq:nusa}) through (\ref{eq:nuc})),
we get: $A_*\sim 1.7$, $\epsilon_B\sim1.1\times 10^{-5}$ and $\epsilon_e \sim
2.0\times 10^{-3}$ and $E\sim 1.6\times 10^{54} \erg$.  These values for the
GRB parameters are in similar ranges as results from previous modelings
\citep{panaitescu01}.  The value of $\epsilon_B$ is somewhat unusual, but not
out of the ordinary \citep[see e.g.][]{Kumar+10external}, and is consistent
with no magnetic field enhancement in the forward shock, beyond the usual shock
compression. Though values have to be considered with precaution as they heavily
depend on the parameter $p$, they indicate that the inverse Compton cooling
might be important, as suggested by \citet{Tang+14ic}.  Indeed, the power in
the synchrotron self-Compton component is $Y$ times the power in the
synchrotron component with $Y$ the Compton parameter, which for
$\epsilon_e>\epsilon_B$ is $Y\approx\sqrt{\epsilon_e/\epsilon_B}\sim13$.

{ Equations (\ref{eq:nusa}) through (\ref{eq:nuc}) suffer from different types of
uncertainties. The value of $\nu_m$ is an upper limit, $\nu_c$ is constrained
within a factor of $\sim$ few to the high energy limit of XRT. Moreover, the
exponents in Equations (\ref{eq:nusa}) and  (\ref{eq:fnusa}) have a strong
dependence on the value of the parameter $p$. For these reasons, it is difficult
to estimate the uncertainties for the derived parameters, and they should be
treated as approximate.  }

\subsection{Forward-plus-reverse shock scenario}
From the radio data alone, there is no strong evidence for a reverse-shock
component in GRB\,130907A.  However, the radio spectra and light curves of
GRB\,130907a are somewhat reminiscent of the very bright GRB\,130427a
\citep[e.g.,][]{Perley2014},  in which the radio data have been interpreted as
a combination of a reverse and a forward-shock, or a two component jet.
Moreover, the 5\,GHz emission  of GRB 130907a seems to be somewhat better
described by a forward+reverse shock model (see Figure \ref{fig:radiolc}).
Lastly, assuming such a component is present would alleviate the stringent
constraints on the underlying physical parameters. E.g.  $\nu_m\lesssim
20\GHz\lesssim \nu_{\rm SA}$ at t=0.19\,d can be realized with less extreme
parameters (e.g.  $\epsilon_e$) if we consider a reverse shock.  Instead of
Equation \ref{eq:num}, we have $\nu_m^{\rm RS} = 1.1 \GHz ~ E_{54}^{6/7}
A_{*,-1}^{-5/14} \epsilon_{e,-1}^2 \epsilon_{B,-2}^{1/2} \Gamma_2^{24/7}
(t/0.19\,d)^{-13/7} \lesssim 20\GHz$.  Here, we assume a thin shell case for
the reverse shock and $\Gamma$ is the coasting Lorentz factor of the burst.
This requirement is obviously less demanding than  Equation \ref{eq:num}, and
since the strongest dependence is on the $\epsilon_e$ parameter (disregarding
the Lorentz factor which does not enter in the forward shock calculations),
ascribing the initial rise to the reverse shock results in less extreme
$\epsilon_e$.  

The forward-reverse shock scenario introduces new parameters (e.g. temporal
slope of the decaying reverse shock component and the rising slope and peak of
the forward shock) which we are unable to constrain: in Figure
\ref{fig:radiolc} we plot a putative reverse-plus-forward-shock model
(dot-dashed and dashed curves).  We assumed an early reverse-shock component
with temporal slopes fixed from theory \citep[thin shell case in an ISM medium:
$ \alpha=\{-9/4, (27p+7)/35 \sim 2.38 \} $; e.g.,][]{Gao+13syn}. For the forward
shock, we fix the rising slope index to $\alpha=-5/4$, the peak time to 1.5\,d
and fit for the decaying slope.  We use the ISM case here, because the only
radio data in the wind regime (according to our model) is the Epoch I point.
For the wind case (before the termination shock), one has steeper rising
temporal indices for the reverse- ($\alpha=-65/42$) and forward-shocks
($\alpha=-7/4$). These will not change the overall properties of the
components.

We apply both the forward-shock only and reverse-plus-forward shock scenarios
to the 15\,GHz light curve.  Both models give a good description of the data.
In an attempt to discriminate between the two, we transform the models to
5\,GHz and 8.5\,GHz to compare with observations.  While the 8.5 GHz data
appears to be better described by the forward-shock-only model, the 5 GHz
measurements agree more with the two-component scenario (see Figure
\ref{fig:radiolc}). The first data point at 5\,GHz is over-predicted by both
scenarios and it appears problematic for the forward-shock-only model as it is
$\sim 5$ standard deviations from this model.  

{ In conclusion, within the limitations of the presented dataset, we favor the
forward-shock only model (see previous Section) when compared to a
forward+reverse shock model because the former is simpler and gives a similarly
good description of the data. Being simpler, the forward shock model also
allows us to solve for the microphysical parameters (whereas the
forward+reverse shock model does not). However, we stress that our case study
for this burst clearly indicates that a good temporal coverage at the highest
radio frequencies is necessary to securely identify salient features of the
reverse shock. }

\section{Conclusion}
\label{sec:conclusion}
We have presented early radio detections of GRB\,130907A with the VLA and
subsequent observations as late as 42 days after the burst.  Early radio
observations are important for identifying potential reverse shock signatures.
We complemented our radio observations with freely available data from the
literature.

GRB\,130907A is unusual in two respects: a chromatic steepening of
the X-ray light curve at 0.23\,d since burst which is not commonly observed,
and an early-time X-ray slope which is hard to reconcile with the observed
radio decay.  It is a unique burst in that it has very early self-absorbed
radio observations and measured spectra spanning from radio to the GeV range.

To accommodate these features, we constructed a model where a blast wave
propagates initially into a wind medium, then transitions into a constant
density ISM. A simple forward-shock synchrotron model explains almost all of
the features.  We also considered a model where both reverse and forward shocks
are present, but find no definitive evidence to prefer this to the simpler
forward-shock-only scenario. { We note however that the reverse and
forward-shock scenario also provides a good fit, but the data is  not
constraining enough to argue for or against the inclusion of the reverse
shock.} In order to account for the spectral evolution in the X-rays, we
suggest a cooling break passing in band as the shock enters the ISM.  

From all the observational constraints, { within the framework of the
forward shock only model,} we derive the relevant physical parameters for this
burst. The normalization of the wind density profile is $A_*=1.7$ in units of
mass loss over wind speed. The magnetic field parameter is $\epsilon_B \sim
1.1\times 10^{-5}$,  while the parameter for the energy in electrons is
$\epsilon_e \sim 2.0\times 10^{-3}$. The total kinetic energy of this burst is
comparable to the energy released in $\gamma$-rays, $E\sim 1.6\times 10^{54}
\erg$. These parameters have a strong dependence on the power-law index of the
electron distribution, which we set to $p\approx2.38$.

With the isotropic equivalent energy in excess of $10^{54}\erg$,  and
beaming corrected energy in excess of $3\times10^{52}$ erg, GRB\, 130907A is
part of the hyper-energetic bursts. { If GRB\,130907 has a double-jet
structure, the beaming corrected energy of the narrow component reduces to the
more typical value of $10^{51}$ erg.}  The properties of this burst can be
compared to the broader sample of LAT-detected GRBs or bursts with an
identified wind-ISM transition.

Comparing our derived parameters with those of the LAT-detected  sample
\citep[see Table 3 in][]{Racusin+11fermiswift}, we find this burst typical in
many respects. In terms of both kinetic ($1.6\times10^{54}\erg$) and radiated
($3\times10^{54}\erg$)  isotropic-equivalent energy, this is an average GRB.
Furthermore, in terms of energy conversion efficiency
($E_{\gamma}/(E_\gamma+E_K)\sim 65\%$), this burst is in the middle of the
distribution of LAT-detected GRBs.

The value of the termination shock radius ($\sim 7.7\times 10^{17} \cm$) is
consistent with the ones derived from similar wind-ISM transition modeling.
However, they all suffer from the inconsistency already noted in the literature
\citep[e.g.][]{Jin+09bubble} between numerical models of Wolf-Rayet (WR) stars'
termination shock radii ($\gtrsim 3\times 10^{18} \cm$) and GRB-deduced
values. This can be mitigated e.g. if the stellar wind is anisotropic
\citep{Eldridge07aniso,vanMarle+08aniso}, if the WR wind is weak, or if it
resides in a high density or high pressure ISM \citep{vanMarle+06termrad}.

The parameters obtained for GRB\,130907A, perhaps with the exception of the
density parameter, are in the same range among the GRBs with claimed wind-ISM
transition. The $A_*$ parameter of this burst is larger by $\sim 2$ orders of
magnitude compared to other GRBs. For similar total isotropic energy ($\sim
10^{54} \erg$) GRB\,081109 \citep{Jin+09bubble} has $\epsilon_e$ and
$\epsilon_B$ parameters larger by an order of magnitude, while the wind density
parameter ($A_*$) is two orders of magnitude lower. GRB\,081109 has a low
efficiency ($\sim 1\%$) compared to GRB\,130907A ($65\%$). In case of
GRB\,050319 \citep{Kamble+07transition}, $\epsilon_e$ is the same order of
magnitude as in our case, $\epsilon_B$ is significantly (3-4 orders of
magnitude) higher, and $A_*$ is 2-3 orders of magnitude lower.  

In conclusion, GRB\,130907A poses intriguing challenges in the modeling of its
multi-wavelength observations, since a simple ISM or wind density profiles fail to
account for the entire set of observations. Invoking the wind-ISM transition is
a natural extension of the wind-only scenario, and one would expect to have such
a transition in all wind density profiles as the shock reaches the ISM surrounding 
the progenitor star. This burst adds to the small number of
GRBs showing this transition.  \\

{\it Acknowledgement - } We thank Phil Evans and Bing Zhang  for valuable
discussions, and the anonymous Referee for useful comments.  PV acknowledges
support from Fermi grant NNM11AA01A and OTKA NN 111016 grant. AC acknowledges
partial support from the NASA-\textit{Swift} GI program via grants
13-SWIFT13-0030 and 14-SWIFT14-0024. Support for SBC was provided by NASA
through the Fermi grant NNH13ZDA001N.  This work made use of data supplied by
the UK Swift Science Data Centre at the University of Leicester.

\bibliographystyle{hapj}




\end{document}